\documentstyle[12pt,prc,aps,epsfig]{revtex}
\tighten
\begin{document}
\draft
\title{Chiral $NN$ model and $A_y$ puzzle}
\author{D. R. Entem\footnote{On leave from University of Salamanca, Spain.
Electronic address: dentem@uidaho.edu} 
and R. Machleidt\footnote{Electronic address: machleid@uidaho.edu}}
\address{Department of Physics, University of Idaho, Moscow,
Idaho 83844}
\author{H. Wita\l a\footnote{Electronic address: witala@if.uj.edu.pl}}
\address{Institute of Physics, Jagiellonian University, PL-30059 Cracow,
Poland}

\date{\today}

\maketitle

\begin{abstract}
We analyze
the results by chiral $NN$ models for the two-nucleon
system and calculate the predictions
for the nucleon vector analyzing power of elastic
nucleon-deuteron ($Nd$) scattering, $A_y$, by these models.
Our conclusion is that
a {\it quantitative} chiral two-nucleon potential 
does not resolve the $Nd$ $A_y$ puzzle (when only two-body forces
are included).
\end{abstract}



The term {\it $A_y$ puzzle} refers to the inability to explain
the nucleon vector analyzing power $A_y$ in elastic 
nucleon-deuteron ($Nd$) scattering
below 30 MeV laboratory energy for the incident nucleon
by means of three-body calculations in which only two-nucleon
forces are applied.
The problem showed up as soon as it was possible to conduct
three-body continuum calculations
{\it with realistic $NN$ potentials}.
The first such calculation was performed by 
Stolk and Tjon~\cite{ST78}
in 1978 using the Reid soft-core potential~\cite{Rei68}, and
the first calculations with (a separable representation
of) the Paris potential~\cite{Lac80} were conducted 
by the Graz-Osaka group in 1987~\cite{PH87};
in both cases, the $A_y$ predictions showed the characteristic problem.
Finally, the `puzzle' became proverbial
when rigorous three-nucleon continuum Faddeev calculations
using realistic forces were started on a large scale~\cite{WGC87}.
Over the years, many measurements and calculations
of $Nd$ $A_y$ were performed (including the $pd$ reaction
that involves the Coulomb force~\cite{KVR95})
which all confirmed
that the problem was real (see Ref.~\cite{Glo96} for a review):
For energies below 20 MeV, the $A_y$ is predicted about
30\% too small in the angular region around 120 deg center-of-mass
angle where the maximum occurs.

There have been many attempts to solve the problem.
Already in the very early stages of three-body continuum calculations,
when only schematic $NN$ potentials were applied, it was noticed
that the $Nd$ $A_y$ predictions
depend very sensitively on the strength of the input $NN$ potential
in the triplet $P$ waves~\cite{Pie72,Dol72}---a sensitivity that 
was confirmed in later calculations using realistic forces~\cite{WGC89}.
Based upon this experience,
Wita{\l}a and Gl\"ockle~\cite{WG91} 
showed in 1991 that 
small changes
in those $^3P$ wave potentials 
could remove the discrepancy.
This finding gave rise to systematic investigations of the question
whether the small variations of the low-energy phase shifts of, 
particularly, those triplet $P$ waves necessary to explain
the $Nd$ $A_y$ are consistent with the $NN$ data base.
While Tornow and coworkers~\cite{Tor98}
suggest that the low-energy $NN$ data may leave some
lattitude in the $NN$ $^3P$ waves that could improve the predictions
for $Nd$ $A_y$,
H\"uber and Friar~\cite{HF98} find that it is not possible
with reasonable changes in the $NN$ potential to increase the $Nd$ $A_y$
and at the same time to keep the two-body observables unchanged.

Another important observation has been that
conventional three-nucleon forces 
(when added to a realistic two-nucleon potential)
change the predictions for $Nd$ $A_y$ only slightly
and do not improve them~\cite{HWG93,KVR95}.
Therefore,
the general perception in the community has shifted towards the
believe that the $A_y$ puzzle is the `smoking gun' for 
new types of three-nucleon forces
\cite{FHK99,Hub01,Fri01,CS01}
or new physics~\cite{SWM00}.

However, very recently, there has been an apparent indication
that the above conclusion/believe may be premature.
It was reported~\cite{Epe01} that with a two-nucleon force of a new type,
namely, one that is based upon chiral effective field theory, 
the $A_y$ puzzle is resolved.

In recent years, effective field theory methods have become
increasingly popular in nuclear physics. 
The reason for this development is the need to link
conventional nuclear physics methods one way or the other
to the underlying theory of strong interactions, QCD.
After quark cluster models had only a limited success,
it was recognized that the symmetries of QCD are more important
than the high-energy degrees of freedom of QCD (quarks and gluons).
The effective field theory concept distinguishes between different
energy scales and assignes appropriate degrees of freedom
for each scale while observing the over-all symmetries.
For traditional nuclear physics with energies below 1 GeV,
the right degrees of freedom are nucleons and pions
interacting via a force that is controlled by (broken)
chiral symmetry.

The derivation of the nuclear force from chiral effective field
theory was initiated by Weinberg \cite{Wei90} and pioneered
by Ord\'o\~nez \cite{OK92} and van Kolck \cite{ORK94,Kol99}.
Subsequently, many researchers became interested in the subject
\cite{CPS92,RR94,KBW97,KGW98,Kai99,EGM98,KSW96,FST97,Par98,Coh97,RS99,Bea01}.
As a result, efficient methods for deriving
the nuclear force from chiral Lagrangians emerged and
the quantitative nature of the chiral $NN$ potential
improved. This trend shows up, in particular, 
in the excellent work by Epelbaum {\it et al.}~\cite{EGM98} where the chiral 
$NN$ force was constructed using a unitary transformation and applying 
systematic power counting in next-to-leading order (NLO) and NNLO;
and it is this potential in NLO that was applied in Ref.~\cite{Epe01},
resulting---seemingly---in a resolution of the
long-standing $A_y$ puzzle. 
%
It is the purpose of this note to critically investigate
the predictions by the chiral $NN$ model and the implications 
for the $Nd$ $A_y$ puzzle.

We start our investigation by taking a close look at 
important phase shifts of two-nucleon scattering.
In Table I, we list $S$-wave phase shifts 
and, in Table~II, we show $^3P$-wave phase shifts 
for energies between 1 and 200 MeV.
Since charge-dependence of the $NN$ interaction is
not a crucial factor in the $A_y$ puzzle~\cite{Tor98}, 
and since the present chiral $NN$ potentials are all 
charge-independent and adjusted
to the neutron-proton ($np$) data, 
we consider $np$ phase shifts and $np$ data.

It is of interest to compare the phase shifts
produced by the chiral NLO (next-to-leading order) model by
Epelbaum {\it et al.}~\cite{EGM98} (which is the chiral potential applied in
Ref.~\cite{Epe01}) with the empirical ones from the Nijmegen
multi-energy $np$ analysis~\cite{Sto93} (PWA93) and the predictions by
one representative of the family of the high-precision potentials
constructed in the 1990's (CD-Bonn~\cite{MSS96,Mac01}).
In $S$ waves (Table~I), there is, generally, good agreement up to 50 MeV.
Above 50 MeV, differences between NLO and PWA93 show up and
increase with energy. However, since the $S$-waves are not very
important for $Nd$ $A_y$, this may not have much impact on the
predictions.

We turn now to the triplet-$P$ waves (Table~II) which are
crucial for $Nd$ $A_y$, and focus, first, on the energy range below 30 MeV.
The NLO $^3P_0$ phase shifts are about 2\% larger than PWA93, 
which is not significant.
For $^3P_1$, the
differences are more drastic: The NLO value at 10 MeV is about
5\% smaller and the one at 25 MeV is 3\% smaller than PWA93.
Finally, the NLO prediction for $^3P_2$ at 25 MeV
is enhanced by 13\%.

To understand how variations of the $^3P$ phase shifts may effect
observables, we consider
the spin-orbit phase shift combination,
\begin{equation}
\Delta_{\rm LS} = \frac{1}{12} \left( -2\, \delta_{^3P_0} -3\, \delta_{^3P_1}
             +5\, \delta_{^3P_2} \right)
\, ,
\end{equation}
which is a measure for the strength of the spin-orbit
force. Results are shown in Table~III.
At 10 MeV, one obtains 
$\Delta_{\rm LS}^{\rm PWA93}=0.203$ deg 
and 
$\Delta_{\rm LS}^{\rm NLO}=0.212$ deg 
for PWA93 and NLO, respectively, implying that the NLO value
is larger by 4.4\% as compared to PWA93.
At 25 MeV, the corresponding figures are
$\Delta_{\rm LS}^{\rm PWA93}=0.93$ deg 
and 
$\Delta_{\rm LS}^{\rm NLO}
=1.09$ deg implying that NLO is 17\% larger.
These numbers show that the spin-orbit force of NLO is enhanced
as compared to the Nijmegen analysis and similar enhancements are
obtained when comparing to any of the high-precision potentials,
like CD-Bonn (cf.\ Table~III).
It is well-known that the two-nucleon spin-orbit force
is magnified in the $Nd$ system. Therefore, 
a moderate enhancement of the $NN$ $LS$-potential leads to a
substantial enlargement of the $Nd$ $A_y$.
Thus, the enhanced NLO spin-orbit force as reflected in the larger 
values for $\Delta_{\rm LS}^{\rm NLO}$
could very well be the explanation of
the large $Nd$ $A_y$ predictions by NLO reported in Ref.~\cite{Epe01}.

Next, we consider the $^3P$ phase shifts above 30 MeV.
Here, the differences between PWA93 and CD-Bonn, on the one hand,
and NLO, on the other, are in general larger and increase with energy (cf.\ Table~II).
This trend is most dramatic in $^3P_2$ where the discrepancies quickly
grow into the hundreds of standard deviations.
The sign of these differences are such as to drastically
enhance the spin-orbit force, see Table~III for the value at 50 MeV.
Note that the $NN$ $t$-matrix, on- and off-shell, is input 
to the three-body continuum calculations.
Thus, the description of the two-nucleon data at energies above
30 MeV has impact on low-energy three-body predictions.
Therefore, the unrealistically strong NLO spin-orbit force above 30 MeV
may---by means of an off-shell effect---further enhance
the $Nd$ $A_y$ predictions.
However, one may not have much confidence in this type of off-shell effect.

The above observations trigger the question if chiral models can also make
more accurate predictions for $NN$ phase shifts; and if so, what are the
implications for the $Nd$ $A_y$ problem, in such a case.
The natural way is to include higher order terms in power counting 
which should improve not only the quality of the $NN$ phase shift reproduction 
but also extend the energy range in which it works. 
For that purpose, we pick up
the chiral $NN$ potential of Ref.~\cite{EM01} 
(subsequently denoted by `Idaho')
that was recently developed.
In the chiral Idaho model~\cite{EM01}, 
contact terms up to order four are included
which introduces more parameters allowing for a better fit of
the lower partial waves in a much wider energy range.
In Table I and II, it is clearly seen that
the chiral Idaho $NN$ potential reproduces the empirical
$np$ phase shifts of PWA93 up to 200 MeV, accurately.

We consider now the observable $Nd$ $A_y$ which is the focus
of this paper.
We have calculated the predictions for $Nd$ $A_y$ 
at energies 3, 10, and 65 MeV for the incident nucleon.
The results are shown in Fig.~1. 
In this figure, 
the shaded band represents the prediction by the family
of high-precision potentials (using always the $np$
version of those models), namely, CD-Bonn~\cite{Mac01},
Argonne $V_{18}$~\cite{WSS95}, and the Nijmegen potentials
Nijm-I, Nijm-II, and Nijm93~\cite{Sto94}.
The dashed line is the predicition by  the
Idaho chiral $NN$ potential~\cite{EM01} and
it is clearly seen that this prediction 
follows accurately the narrow band made up from the variations among
those high-precision potentials.
In conclusion, at 3 and 10 MeV, we are still having an $A_y$ problem
if the chiral $NN$ potential is a {\it quantitative} one.  

The evidence presented may be perceived as a convincing
proof that a {\it quantitative} chiral potential does not
resolve the $Nd$ $A_y$ puzzle.
However, there remains one objection that can be
raised. In the literature, notably in Ref.~\cite{Tor98},
one can find the suggestion that the $^3P$ waves at low
energy are not as well determined as claimed in PWA93~\cite{Sto93}.
If true, then moderate variations of the $^3P$ phase shifts
at low energy could be consistent with the low-energy $NN$ data.
This variations could be such as to enhance the low-energy spin-orbit force
and, thus, lead to an improved prediction for $Nd$ $A_y$.
For the purpose of seriously checking out this possibility,
we have constructed a variation of the Idaho chiral potential
with modified $^3P$ phase shifts at low energy.
The column `Modified' of Table~II shows the $^3P$ phase shifts
and the corresponding column of Table~III reveals that 
for this fit the spin-orbit force is enhanced, similarly to NLO.
However, in contrast to NLO, the `modified' model is much
more realistic, since the phase shifts do not diverge to
unrealistic values at higher energies.

We have calculated the $Nd$ $A_y$ as predicted by the `modified'
chiral model and find, indeed, a considerable improvement
(see dotted curve in Fig.~1). Is this
the resolution of the $A_y$ puzzle by a chiral $NN$ potential?
To answer this question one needs to know if the modified chiral
model is a realistic and quantitative $NN$ potential.
A precise and reliable answer cannot be given by just looking at
phase shifts. As stressed repeatedly by the Nijmegen group
in the past, only a direct confrontation with the $NN$ data
can reveal if a $NN$ potential is quantitative or not.
For that reason we have calculated the $\chi^2$/datum for the
fit of the world $np$ data as represented by the 
1999 database~\cite{data99}, see Table IV.
For a proper interpretation of the results of Table~IV,
it is necessary to establish a standard concerning what
$\chi^2$/datum represents a `quantitative' reproduction of the
data. This issue was debated a lot in the 1990's, and the
consensus that emerged was that only values below 1.1
are acceptable. Deliberatly, we losen this standard
and consider a fit with $\chi^2/{\rm datum} \leq 1.2$
as `quantitative', while we will perceive higher
values as not acceptable.

Applying this standard, the modified model produces unacceptable
values for $\chi^2$/datum for all energy intervals below 75 MeV
(cf.\ Table~IV). Thus, the modified chiral model is {\it not}
a quantitative one and, consequently, it is {\it not}
the resolution of the $Nd$ $A_y$ puzzle.
Since it is well known that the off-shell character of the $NN$ potential
plays essentially no role
in three-nucleon scattering, one can further draw 
the more general conclusion: No model that reproduces the $NN$ data
correctly can solve that $Nd$ $A_y$ puzzle.

Table~IV shows also the $\chi^2$/datum of the other models discussed
in this paper. It is seen that PWA93, Idaho, and CD-Bonn reproduce
the $np$ data below 210 MeV with the perfect $\chi^2/{\rm datum} = 0.97$
and 0.98. The chiral NLO potential by Epelbaum {\it et al.}~\cite{EGM98}
produces $\chi^2/{\rm datum} = 37$ which is grossly unacceptable.
In fact, only for the interval 0-8 MeV is NLO acceptable.
This range of validity is so tiny that no serious implications can
be drawn from any prediction by this potential.

Finally, we like to take this opportunity to also present an
overview of other interesting $Nd$ observables, which are
shown in Figs.~2-8.
Concerning the deuteron vector analyzing power, $iT_{11}$, at 3 and 10 MeV (Fig.~2), 
it should be noted that a seemingly drastic reduction of the discrepancy 
between theory and $pd$ data seen at 3 MeV has its origin in large effects 
of the long range Coulomb force acting between two protons, which is not 
taken into account in our calculations~\cite{KVR95}. Taking this Coulomb 
force into account, $iT_{11}$ is also  underpredicted in the peak
region~\cite{KVR95}---an equally well-known problem, which is why it would be
appropriate to speak more generally of the vector analyzing 
power puzzle in elastic $Nd$ scattering. 
In almost all cases, the Idaho chiral $NN$ potential follows
the trend of the high-precision potentials.
The only exceptions are $T_{20}$ and $T_{21}$ at 65 MeV where
the chiral potential predictions describe  slightly better the data in the
minimum region  as compared to conventional potentials.
But, apart from this,
the quantitative chiral $NN$ model containing contributions of higher orders 
in power counting does not produce any new signatures.

In summary, our main conclusions are:
\begin{itemize}
\item
A {\it quantitative} chiral two-nucleon potential 
does not resolve the $Nd$ $A_y$ puzzle on the two-body force
level.
\item
Low-energy $^3P_J$ $NN$ phase shifts that ``solve''
the $Nd$ $A_y$ puzzle are inconsistent with the low-energy
$NN$ data.
\end{itemize}
And, finally, as a consequence of the above two points, one may
expect that
{\it no quantitative two-nucleon force---no matter what the
basis is, pure phenomenology, meson theory, chiral EFT, 
or anything---will ever solve the $Nd$ $A_y$ puzzle.}

An accurate $NN$ model requires to take
chiral perturbation theory ($\chi$PT) to order four.
At that order, also many three-body force (3NF) terms occur.
According to the basic rules of $\chi$PT,
all two- and many-body terms must be included
for a complete calculation.
Conventional 3NFs were shown to be ineffective for the $A_y$ problem
\cite{HWG93,KVR95}.
The advantage of $\chi$PT is that it provides a systematic
scheme to generate all terms at a given order.
As it turns out, there are several such chiral 3NF terms that were never
considered in few-nucleon physics before
\cite{FHK99,Hub01,Fri01,CS01}.
It is natural to expect the resolution of the $Nd$ $A_y$ puzzle
from such new chiral 3NFs which, therefore, should be at the focus
of future work in the field.

At next-to-leading order (NLO) in $\chi$PT, there are no 3NF contributions. 
So, a calculation with a NLO two-nucleon potential and no 3NF
seems to have formal validity.
However, since at NLO the $NN$ data can only be reproduced 
for $T_{\rm lab} \leq 8$ MeV, such a
calculation is doomed to be inconclusive,
from the outset, and higher order terms must be taken into account
for any meaningful calculation.

\acknowledgements
This work was supported in part by the U.S. National Science
Foundation under Grant No.~PHY-0099444, by the Ram\'on Areces
Foundation (Spain), and by the U.S. Department of Energy, Office of High 
Energy and Nuclear Physics, under Grant No. DE-FG02-97ER41033. The numerical 
calculations have been performed on the CRAY T90 of the John von Neumann 
Institute for Computing in J\"ulich, Germany.

\pagebreak

\begin{table}
\caption{
$^1S_0$ $np$ and $^3S_1$ phase shifts (in degrees).
}
\begin{tabular}{cdddd}
 $T_{\rm lab}$ (MeV) 
 & PWA93$^a$
 & NLO$^b$
 & Idaho$^c$
 & CD-Bonn$^d$
\\
\hline 
\multicolumn{5}{c}{\hspace*{-1.5cm}$\bbox{^1S_0}$} \\                                               
    1 & 
     62.068(30) &
     62.05&
     62.03&
     62.09\\
    5 & 
     63.63(8) &
     63.85&
     63.52&
     63.67\\
   10 &  
     59.96(11) &
     60.26&
     59.80&
     60.01\\
   25 &  
     50.90(19) &
     50.92&
     50.66&
     50.93\\
   50 & 
     40.54(28) &
     39.29&
     40.27&
     40.45\\
  100 & 
     26.78(38) &
     21.68&
     26.87&
     26.38\\
  150 & 
     16.94(41) &
      7.26&
     17.56&
     16.32\\
  200 & 
      8.94(39) &
     -5.53&
      9.58&
      8.31\\
\multicolumn{5}{c}{\hspace*{-1.5cm}$\bbox{^3S_1}$} \\                                               
    1 & 
     147.747(10) &
    147.70&
    147.76&
    147.75\\
    5 & 
     118.178(21) &
    118.29&
    118.20&
    118.18\\
   10 &  
     102.611(35) &
    102.84&
    102.64&
    102.62\\
   25 &  
      80.63(7) &
     80.69&
     80.67&
     80.63\\
   50 & 
      62.77(10) &
     61.74&
     62.80&
     62.73\\
  100 & 
      43.23(14) &
     38.65&
     43.13&
     43.06\\
  150 & 
      30.72(14) &
     21.67&
     30.41&
     30.47\\
  200 & 
      21.22(15) &
      7.20&
     20.90&
     20.95\\
\end{tabular}
$^a$Nijmegen multi-energy $np$ analysis~\cite{Sto93}.
Numbers in parentheses give the uncertainties in the last digits.
\\
$^b$Chiral next-to-leading order (NLO) potential 
by Epelbaum {\it et al.}~\cite{EGM98}
using a Gaussian cutoff with cutoff mass $\Lambda = 600$ MeV.\\
$^c$Chiral $NN$ potential by Entem and Machleidt~\cite{EM01}
(`Idaho-B' is used).\\
$^d$Reference~\cite{Mac01}.
\end{table}

\begin{table}
\caption{
Triplet-$P$ $np$ phase shifts (in degrees).
For notation see Table~I.
}
\begin{tabular}{cddddd}
 $T_{\rm lab}$ (MeV) 
 & PWA93$^a$
 & NLO
 & Idaho
 & Modified$^b$
 & CD-Bonn
\\
\hline 
\multicolumn{6}{c}{\hspace*{1.0cm}$\bbox{^3P_0}$} \\                     
    1 &
     0.18  &
      0.19&
      0.18&
  0.18&   
      0.18\\
    5 & 
     1.63(1)  &
      1.67&
      1.64&
  1.61&  
      1.61\\
   10 & 
     3.65(2)  &
      3.72&
      3.69&
  3.62 &    
      3.62\\
   25 & 
     8.13(5)  &
      8.22&
      8.19&
  7.98 &  
      8.10\\
   50 &  
    10.70(9)  &
     10.84&
     10.63&
  10.28 &
     10.74\\
  100 & 
     8.46(11)  &
      8.31&
      8.17&
   7.91 &   
      8.57\\
  150 & 
     3.69(14)  &
      2.52&
      3.60&
   3.66 &  
      3.72\\
  200 & 
    -1.44(17)  &
     -4.10&
     -1.21&
  -0.93 & 
     -1.55\\
\multicolumn{6}{c}{\hspace*{1.0cm}$\bbox{^3P_1}$} \\                                               
    1 &
     -0.11  &
     -0.12&
     -0.11&
    -0.11 &  
     -0.11\\
    5 & 
     -0.94 &
     -0.99&
     -0.93&
    -0.94 & 
     -0.93\\
   10 &
     -2.06 &
     -2.16&
     -2.04&
    -2.09 &
     -2.04\\
   25 & 
     -4.88(1) &
     -5.03&
     -4.82&
    -4.98 &   
     -4.81\\
   50 & 
     -8.25(2) &
     -8.32&
     -8.22&
    -8.57 &  
     -8.18\\
  100 & 
    -13.24(3) &
    -12.66&
    -13.36&
  -13.86 &  
    -13.23\\
  150 & 
    -17.46(5) &
    -15.94&
    -17.59&
  -17.85 & 
    -17.51\\
  200 & 
    -21.30(7) &
    -18.86&
    -21.28&
  -21.18 &
    -21.38\\
\multicolumn{6}{c}{\hspace*{1.0cm}$\bbox{^3P_2}$} \\                                               
    1 &
     0.02  &
      0.02&
      0.02&
    0.02 &   
      0.02\\
    5 & 
     0.25  &
      0.24&
      0.26&
   0.28 &   
      0.26\\
   10 &
     0.71  &
      0.70&
      0.72&
    0.78 & 
      0.72\\
   25 & 
     2.56(1)  &
      2.89&
      2.58&
    2.78 &
      2.60\\
   50 & 
     5.89(2)  &
      8.29&
      5.86&
   6.29 &    
      5.93\\
  100 & 
    10.94(3)  &
     22.61&
     10.77&
   11.24 &  
     11.01\\
  150 & 
    13.84(4)  &
     35.98&
     13.72&
 13.82 &   
     13.98\\
  200 & 
    15.46(5)  &
     44.31&
     15.58&
   15.33 &   
     15.66\\
\end{tabular}
$^a$Uncertainties smaller than 0.005 degrees are not shown.\\
$^b$Modified version of the Idaho potential with enhanced spin-orbit
force at low energies.
\end{table}

\pagebreak

\begin{table}
\caption{
Spin-orbit phase shift combination $\Delta_{\rm LS}$ (in degrees),
Eq.~(1), at 10, 25, and 50 MeV for various models
explained in Table I and II.
}
\begin{tabular}{cddddd}
 $T_{\rm lab}$ (MeV) 
 & PWA93
 & NLO
 & Idaho
 & Modified
 & CD-Bonn
\\
\hline 
10&0.203&0.212&0.195&0.244&0.207\\
25&0.93&1.09&0.92&1.07&0.94\\
50&2.73&3.73&2.73&3.05&2.73\\
\end{tabular}
\end{table}

\vspace*{2cm}

\begin{table}
\caption{
$\chi^2$/datum for the reproduction of the 1999 $np$ 
database~\protect\cite{data99}
up to $T_{\rm lab} = 210$ MeV
by various models explained in Table~I and II.
}
\begin{tabular}{ccddddd}
 Bin (MeV) 
 & \# of data
 & PWA93
 & NLO
 & Idaho
 & Modified
 & CD-Bonn
\\
\hline 
0-8 & 81 & 1.05 & 1.05 & 1.03 & 1.32 & 1.06 \\
8-17 & 192&0.86&1.22&0.85&1.87&0.91\\
17-35&292&0.82&2.15&0.81&1.27&0.81\\
35-75&340&1.03&10.8&1.09&1.24&1.01\\
75-125&239&1.00&11.6&0.99&1.06&0.97\\
125-183&414&1.06&95.7&1.07&1.15&1.03\\
183-210&141&0.96&111.7&0.97&1.18&1.01\\
\hline
0-210&1699&0.97&37.0&0.98&1.27&0.97
\end{tabular}
\end{table}

\pagebreak

\begin{figure}
\vspace*{0.5cm}
\hspace*{2.0cm}
\epsfig{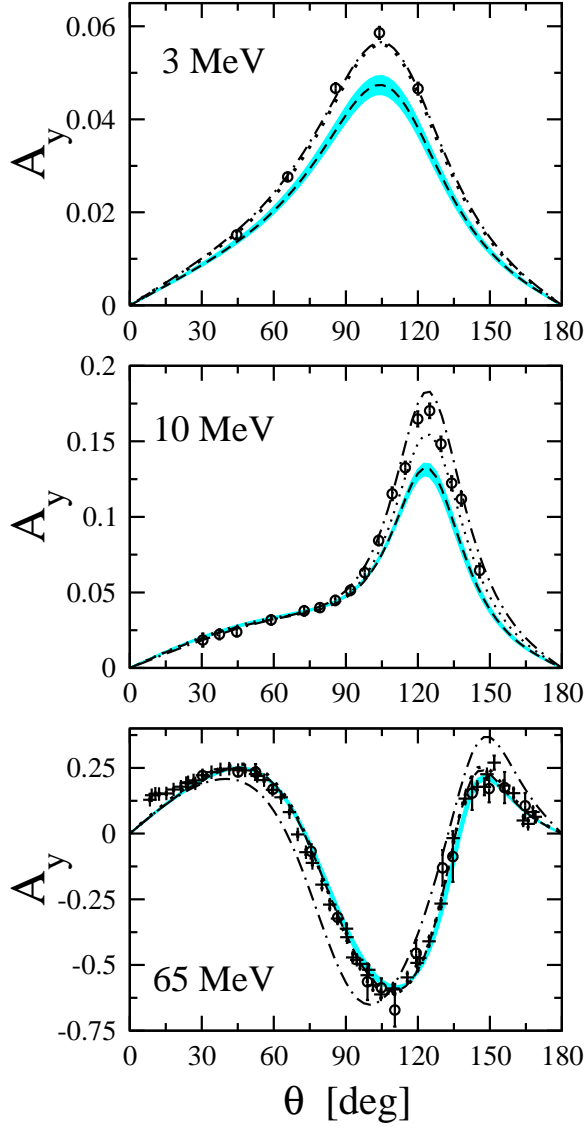}
\vspace*{1.0cm}
\caption{Nucleon analyzing power $A_y$ for elastic $Nd$
scattering at 3, 10, and 65 MeV. 
The shaded band represents the predictions by the 
high-precision potentials 
CD-Bonn~\protect\cite{Mac01},
Argonne $V_{18}$~\protect\cite{WSS95}, and the Nijmegen potentials
Nijm-I, Nijm-II, and Nijm93~\protect\cite{Sto94}
(using always the $np$ versions of these potentials). 
The dashed line is the prediction by  the
Idaho chiral $NN$ potential~\protect\cite{EM01}.
The dotted curve represents the result from the `Modified'
chiral potential (see text) and the dash-dot line is
predicted by the NLO chiral potential by 
Epelbaum {\it et al.}~\protect\cite{EGM98}.
Data at 3 MeV are from~\protect\cite{43} ($nd$, squares), 
at 10 MeV  from~\protect\cite{45} ($nd$, squares), and at 65 MeV 
from~\protect\cite{47} 
($nd$, squares) and 
~\protect\cite{48} ($pd$, crosses).
}
\end{figure}

\pagebreak

\begin{figure}
\vspace*{1.0cm}
\hspace*{2.0cm}
\epsfig{figure=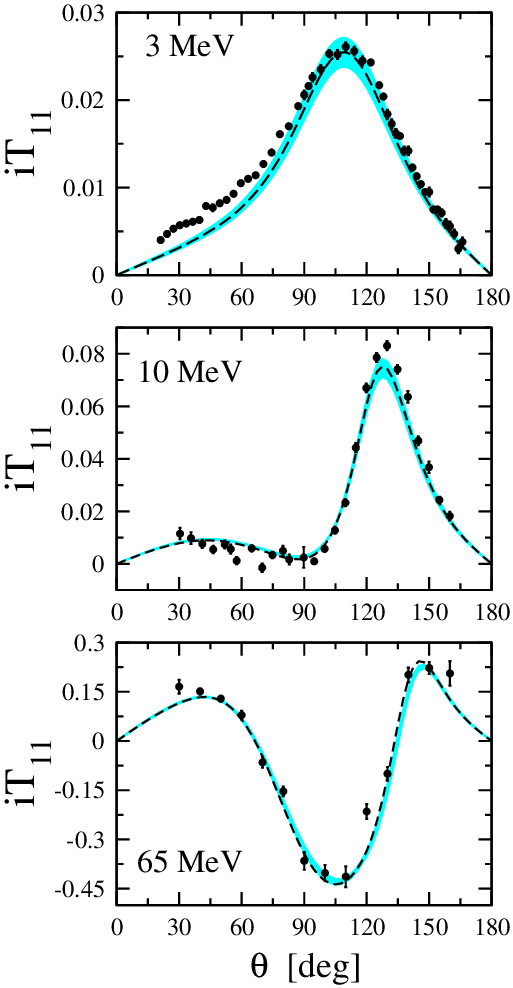,height=17cm}
\vspace*{1.0cm}
\caption{Deuteron vector analyzing power $iT_{11}$
for elastic $Nd$ scattering. Energies, bands, and curves as in Fig.~1.
The circles are $pd$ data: at 3 MeV from ~\protect\cite{44}, 
at 10 MeV from ~\protect\cite{46}, and at 65 MeV from ~\protect\cite{50}.}
\end{figure}

\pagebreak

\begin{figure}
\vspace*{1.0cm}
\hspace*{2.0cm}
\epsfig{figure=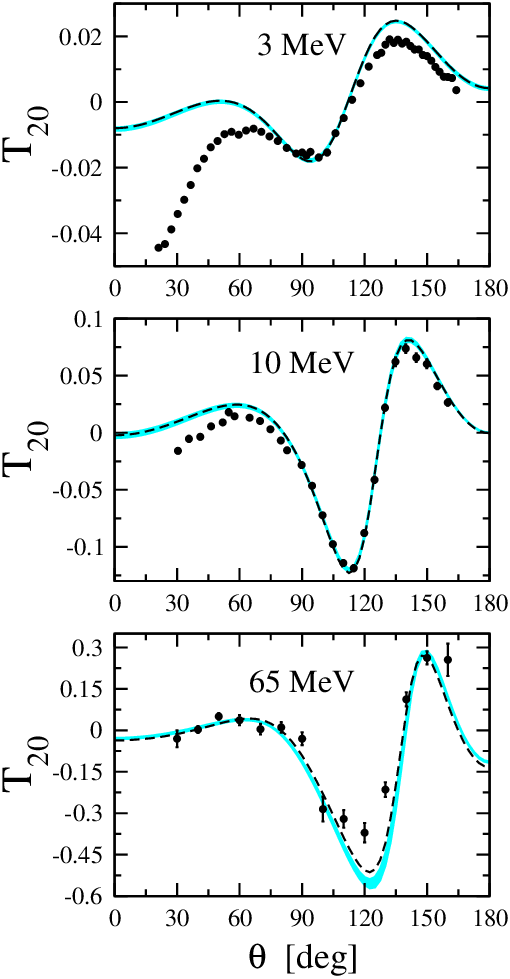,height=17cm}
\vspace*{1.0cm}
\caption{Tensor analyzing power $T_{20}$ for
elastic $Nd$ scattering. Energies, bands, and curves as in Fig.~1.
The circles are $pd$ data: at 3 MeV from ~\protect\cite{44}, 
at 10 MeV from ~\protect\cite{46}, and at 65 MeV from ~\protect\cite{50}.}
\end{figure}

\pagebreak

\begin{figure}
\vspace*{1.0cm}
\hspace*{2.0cm}
\epsfig{figure=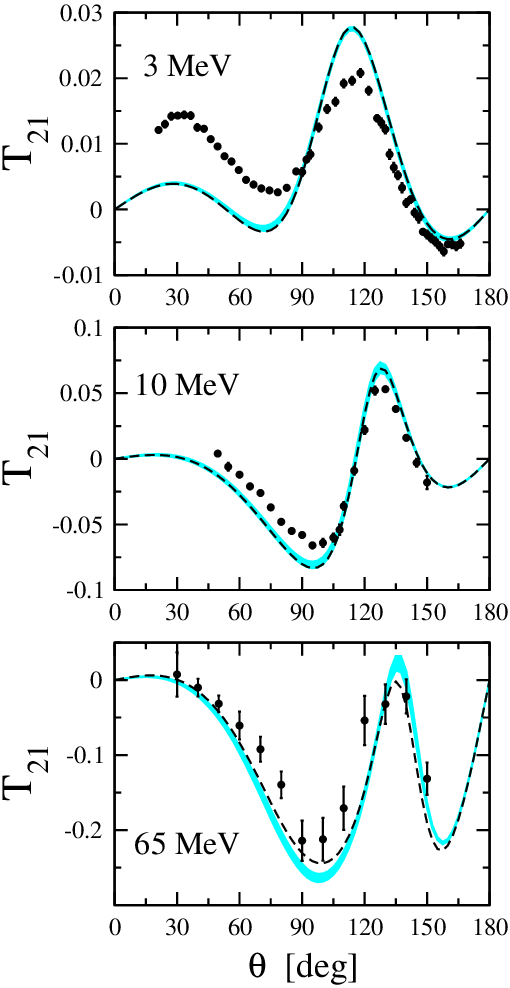,height=17cm}
\vspace*{1.0cm}
\caption{Tensor analyzing power $T_{21}$ for
elastic $Nd$ scattering. Energies, bands, and curves as in Fig.~1.
The circles are $pd$ data: at 3 MeV from ~\protect\cite{44}, 
at 10 MeV from ~\protect\cite{46}, and at 65 MeV from ~\protect\cite{50}.}
\end{figure}

\pagebreak

\begin{figure}
\vspace*{1.0cm}
\hspace*{2.0cm}
\epsfig{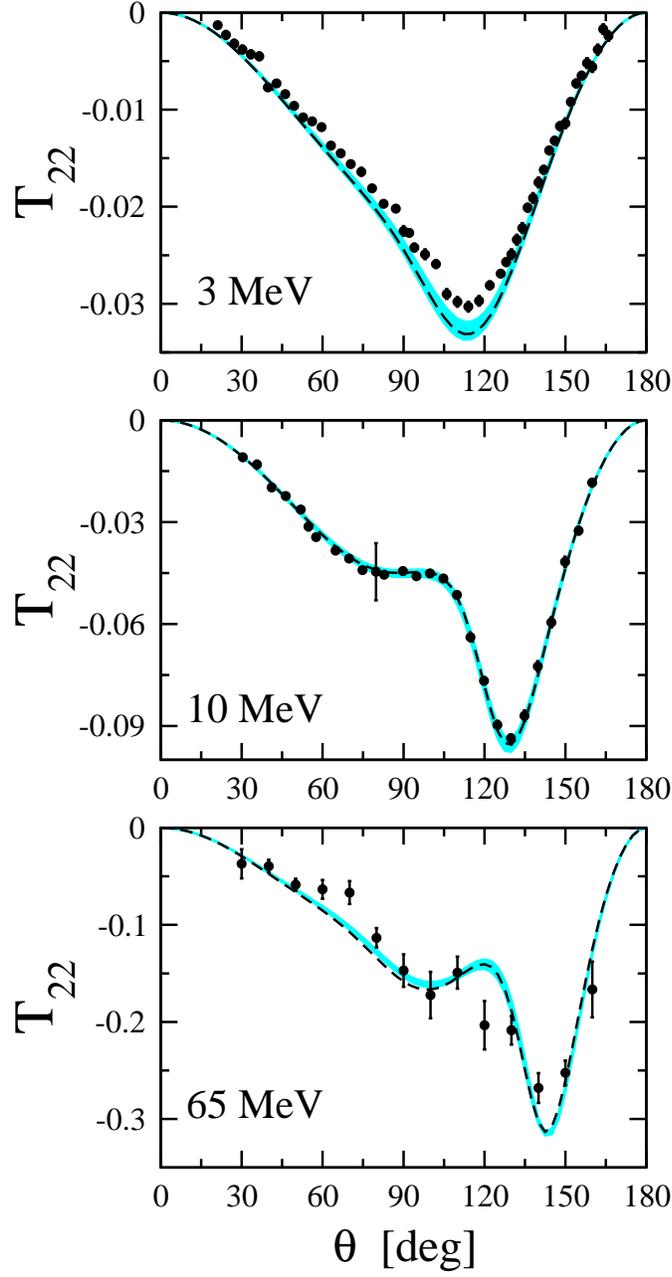}
\vspace*{1.0cm}
\caption{Tensor analyzing power $T_{22}$ for
elastic $Nd$ scattering. Energies, bands, and curves as in Fig.~1.
The circles are $pd$ data: at 3 MeV from ~\protect\cite{44}, 
at 10 MeV from ~\protect\cite{46}, and at 65 MeV from ~\protect\cite{50}.}
\end{figure}

\pagebreak

\begin{figure}
\vspace*{1.0cm}
\hspace*{2.0cm}
\epsfig{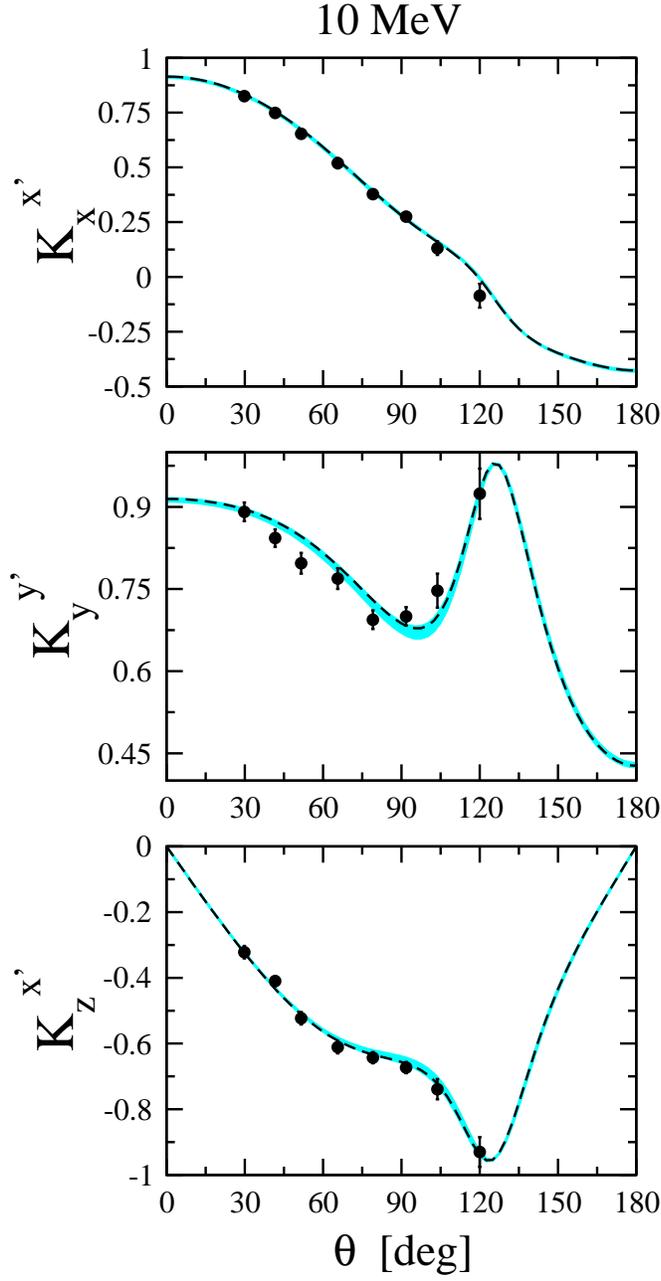}
\vspace*{1.0cm}
\caption{
Spin transfer coefficients
$K_x^{x'}$,
$K_y^{y'}$, and
$K_z^{x'}$
for elastic $Nd$ scattering
at 10 MeV. Bands and curves as in Fig.~1.
The circles are $pd$ data from ~\protect\cite{46}.}
\end{figure}

\pagebreak

\begin{figure}
\vspace*{1.0cm}
\hspace*{2.0cm}
\epsfig{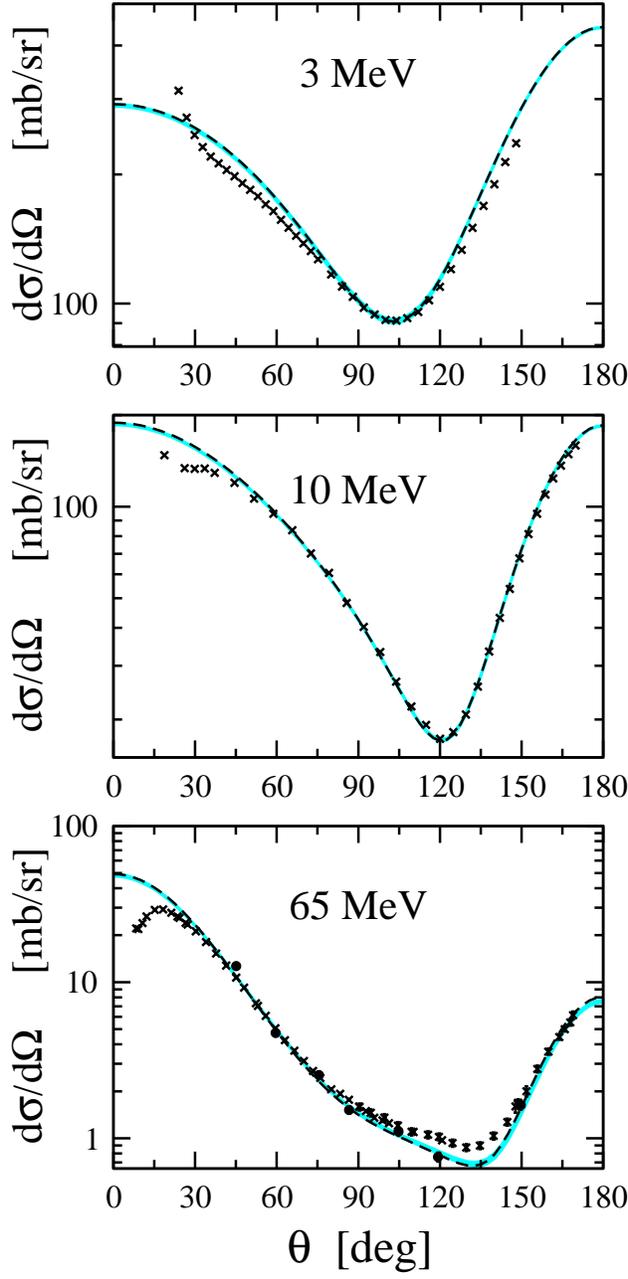}
\vspace*{1.0cm}
\caption{
Differential cross section
$d\sigma/d\Omega$ of elastic $Nd$ scattering.
Energies, bands, and curves as in Fig.~1.
The crosses are $pd$ data: at 3 MeV from ~\protect\cite{49}, 
at 10 MeV from ~\protect\cite{46}, and at 65 MeV from ~\protect\cite{48}. 
The circles at 65 MeV are $nd$ data from ~\protect\cite{47}.}
\end{figure}

\pagebreak

\begin{figure}
\vspace*{1.0cm}
\hspace*{1.0cm}
\epsfig{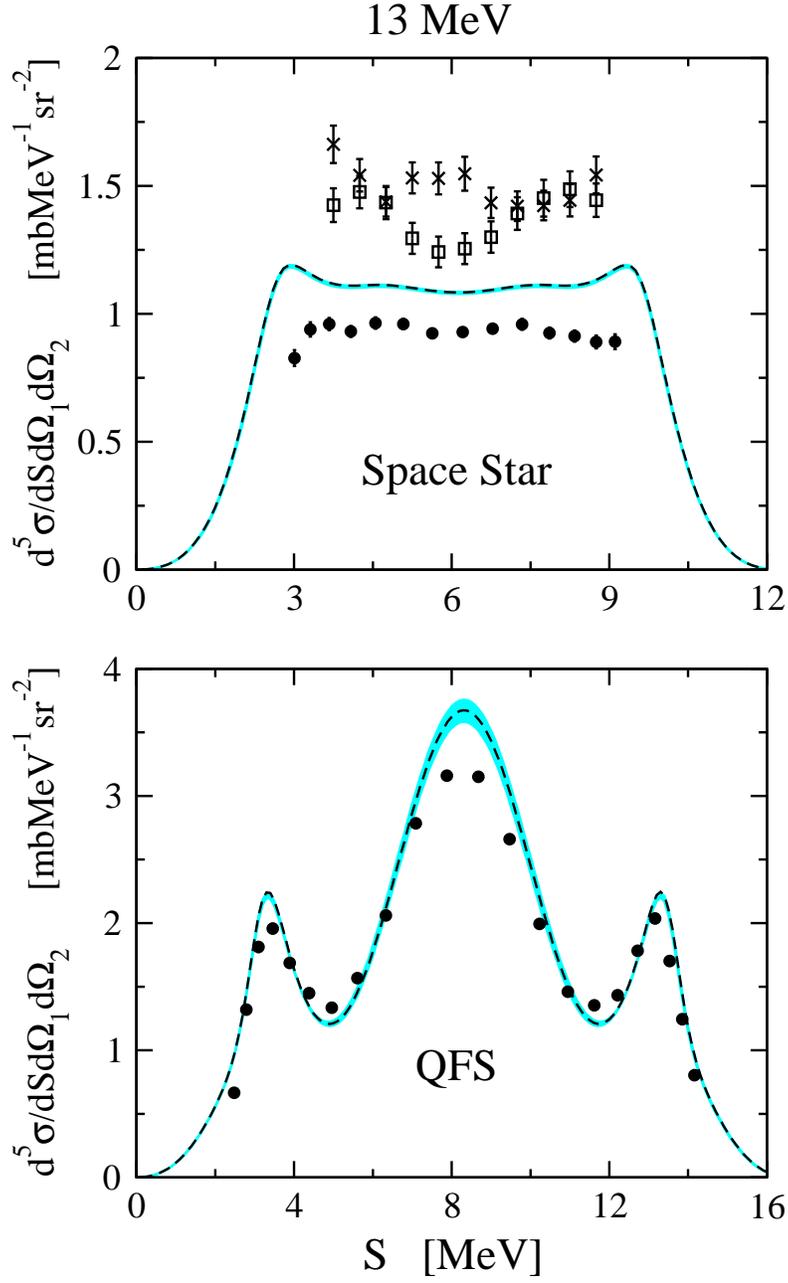}
\vspace*{1.0cm}
\caption{Neutron-deuteron breakup cross sections
for the space-star and quasi-free scattering (QFS)
configurations at 13 MeV along the kinematical locus $S$. 
Bands and curves as in Fig.~1. 
The circles are $pd$ data from ~\protect\cite{51}. 
The crosses and squares are $nd$ data from ~\protect\cite{52} and 
~\protect\cite{53}, respectively.}
\end{figure}


\begin{references}
\bibitem{ST78} C. Stolk and J. A. Tjon, Nucl. Phys. {\bf A295},
384 (1978).
\bibitem{Rei68} R. V. Reid, Ann. Phys. (N.Y.) {\bf 50}, 411 (1968).
\bibitem{Lac80} M. Lacombe, B. Loiseau, J. M. Richard, R. Vinh Mau,
J. C\^{o}t\'{e}, P. Pires, and R. de Tourreil, Phys. Rev. {\bf 21},
861 (1980).
\bibitem{PH87}
W. Plessas and J. Haidenbauer, Few-Body Syst., Suppl. {\bf 
2}, 185 (1987);
Y. Koike and J. Haidenbauer, Nucl. Phys. {\bf A463}, 365c (1987); 
J.Haidenbauer, Y. Koike, W. Plessas and H. Zankel,
Proc.\ Intern.\ Workshop on Few-Body Approaches
to Nuclear Reactions in Tandem and Cyclotron Energy Regions,
edited by S. Oryu and T. Sawada (World Scientific, Singapore, 1987),
p.\ 23.
\bibitem{WGC87} H. Wita\l a, W. Gl\"{o}ckle, and T. Cornelius,
Few-Body Syst., Suppl. {\bf 2}, 555 (1987); 
Nucl. Phys. {\bf A491}, 157 (1989).
\bibitem{KVR95} A. Kievsky, M. Viviani, and S. Rosati,
Nucl. Phys. {\bf A577}, 511 (1994);
Phys. Rev. C {\bf 52}, R15 (1995).
\bibitem{Glo96} W. Gl\"ockle, H. Wita\l a, D. H\"uber, H. Kamada,
and J. Golak, Phys. Rep. {\bf 274}, 107 (1996).
\bibitem{Pie72} S. C. Pieper, Nucl. Phys. {\bf A193}, 529 (1972).
\bibitem{Dol72} P. Doleschall, Phys. Lett. {\bf 38B}, 298 (1972);
Nucl. Phys. {\bf A201}, 264 (1973); {\it ibid.} {\bf A220}, 491 (1974);
P.\ Doleschall, W.\ Gr\"uebler, V.\ K\"onig, P.\ A.\ Schmelzbach,
F.\ Sperisen, and B.\ Jenny, 
Nucl.\ Phys.\ {\bf A380}, 72 (1982).
\bibitem{WGC89} H. Wita\l a, W. Gl\"{o}ckle, and T. Cornelius,
Nucl. Phys. {\bf A496}, 446 (1989).
\bibitem{WG91} H. Witala and W. Gl\"{o}ckle, Nucl. Phys. {\bf A528}, 48 (1991). 
\bibitem{Tor98} 
W. Tornow, H. Wita\l a, and A. Kievsky,
Phys. Rev. C {\bf 57}, 555 (1998);
W. Tornow, and H. Wita\l a,
Nucl. Phys. {\bf A637}, 280 (1998);
T. Tornow and W. Tornow, Few-Body Syst.
{\bf 26}, 1 (1999).
\bibitem{HF98} D. H\"uber and J. L. Friar,
Phys. Rev. C {\bf 58}, 674 (1998).
\bibitem{HWG93} 
D. H\"uber,
H. Wita\l a, 
and W. Gl\"ockle, 
Few-Body Syst. {\bf 14}, 171 (1993);
H. Wita\l a, 
D. H\"uber,
and W. Gl\"ockle, 
Phys. Rev. C {\bf 49}, R14 (1994).
\bibitem{FHK99} 
D. H\"uber,
J. L. Friar,
and U. van Kolck,
Phys. Rev. C {\bf 59}, 53 (1999).
\bibitem{Hub01}
D. H\"uber,
J. L. Friar,
A. Nogga, H. Wita\l a,
and U. van Kolck,
Few-Body Syst. {\bf 30}, 95 (2001).
\bibitem{Fri01}
J. L. Friar, Nucl. Phys. {\bf A684}, 200 (2001).
\bibitem{CS01}
L. Canton and W. Schadow,
Phys. Rev. C {\bf 64}, 031001 (2001).
\bibitem{SWM00} F. Sammarruca, H. Wita\l a, and X. Meng,
Acta Physica Polonica B {\bf 31}, 2039 (2000).
\bibitem{Epe01} 
E. Epelbaum,
H. Kamada,
A. Nogga,
H. Wita\l a, 
W. Gl\"ockle, 
and U.-G. Mei\ss ner,
Phys. Rev. Lett. {\bf 86}, 4787 (2001).
\bibitem{Wei90} S. Weinberg, Phys.\ Lett.\ B {\bf 251}, 288 (1990);
Nucl.\ Phys.\ {\bf B363}, 3 (1991).
\bibitem{OK92}
C. Ord\'o\~nez
and U. van Kolck,
Phys.\ Lett.\ B {\bf 291}, 459 (1992).
\bibitem{ORK94}
C. Ord\'o\~nez,
L. Ray, and U. van Kolck,
Phys.\ Rev.\ Lett.\ {\bf 72}, 1982 (1994);
Phys.\ Rev.\ C {\bf 53}, 2086 (1996).
\bibitem{Kol99} U. van Kolck, Prog.\ Part.\ Nucl.\ Phys.\ {\bf 43}, 337 (1999).
\bibitem{CPS92} L. S. Celenza, A. Pantziris, and C. M. Shakin,
Phys. Rev. C {\bf 46}, 2213 (1992).
\bibitem{RR94}
C. A. da Rocha and M. R. Robilotta, 
Phys.\ Rev.\ C {\bf 49}, 1818 (1994);
{\it ibid.} {\bf 52}, 531 (1995);
M. R. Robilotta, 
Nucl. Phys. {\bf A595}, 171 (1995);
M. R. Robilotta, 
and C. A. da Rocha, 
Nucl. Phys. {\bf A615}, 391 (1997);
J.-L. Ballot,
C. A. da Rocha, 
and M. R. Robilotta, 
Phys. Rev. C {\bf 57}, 1574 (1998).
\bibitem{KBW97} N. Kaiser, R. Brockmann, and W. Weise,
Nucl.\ Phys.\ {\bf A625}, 758 (1997).
\bibitem{KGW98} N. Kaiser, S. Gerstend\"orfer, and W. Weise,
Nucl.\ Phys.\ {\bf A637}, 395 (1998).
\bibitem{Kai99} N. Kaiser, Phys.\ Rev.\ C
{\bf 61}, 014003 (1999);
{\it ibid.\/} {\bf 62}, 024001 (2000);
{\it ibid.\/} {\bf 63}, 044010 (2001);
nucl-th/0107064.
\bibitem{EGM98} E. Epelbaum, W. Gl\"ockle, and U.-G. Mei\ss ner,
Nucl.\ Phys.\ {\bf A637}, 107 (1998); {\it ibid.\/}
{\bf A671}, 295 (2000).
\bibitem{KSW96} 
D. B. Kaplan, M.J. Savage, and M.B. Wise, 
Nucl. Phys. {\bf B478}, 629 (1996);
D. B. Kaplan, Nucl. Phys. {\bf B494}, 471 (1997);
D. B. Kaplan, M.J. Savage, and M.B. Wise, 
Nucl. Phys. {\bf B534}, 329 (1998);
Phys. Lett. {\bf B424}, 390 (1998).
\bibitem{FST97} R. J. Furnstahl, B. D. Serot, and H.-B. Tang,
Nucl. Phys. {\bf A615}, 441 (1997);
J. V. Steel and R. J. Furnstahl, 
{\it ibid.} {\bf A637}, 46 (1998);
R. J. Furnstahl, J. V. Steel, and N. Tirfessa, 
{\it ibid.} {\bf A671}, 396 (2000);
R. J. Furnstahl, H. W. Hammer, and N. Tirfessa, 
{\it ibid.} {\bf A689}, 846 (2001).
\bibitem{Par98} 
T.-S. Park, K. Kubodera, D. P. Min, and M. Rho,
Phys. Rev. C {\bf 58}, 637 (1998);
Nucl. Phys. {\bf A646}, 83 (1999).
\bibitem{Coh97}
T. D. Cohen, Phys. Rev. C {\bf 55}, 67 (1997);
D. R. Phillips and T. D. Cohen, Phys. Lett. {\bf B390}, 7 (1997);
K. A. Scaldeferri, D. R. Phillips, C. W. Kao, and T. D. Cohen, 
Phys. Rev. C {\bf 56}, 679 (1997);
S. R. Beane, T. D. Cohen, and D. R. Phillips, 
Nucl. Phys. {\bf A632}, 445 (1998).
\bibitem{RS99}
G. Rupak and N. Shoresh, nucl-th/9906077;
P. F. Bedaque, H. W. Hammer, and U. van Kolck,
Nucl. Phys. {\bf A676}, 357 (2000);
S. Fleming, Th. Mehen, and I. W. Stewart,
Nucl. Phys. {\bf A677}, 313 (2000);
Phys. Rev. C {\bf 61}, 044005 (2000).
\bibitem{Bea01}
S. R. Bean, P. F. Bedaque, M. J. Savage, and U. van Kolck, 
nucl-th/0104030.
\bibitem{Sto93} V.\ G.\ J.\ Stoks, R.\ A.\ M.\ Klomp, M.\ C.\ M.\ Rentmeester,
and J.\ J.\ de Swart, Phys.\ Rev.\ C {\bf 48}, 792 (1993).
\bibitem{MSS96}
R. Machleidt, F. Sammarruca, and Y. Song, 
Phys.\ Rev.\ C {\bf 53}, 1483 (1996).
\bibitem{Mac01} R. Machleidt,
Phys.\ Rev.\ C {\bf 63}, 024001 (2001).
\bibitem{EM01} D. R. Entem and R. Machleidt, nucl-th/0107057;
nucl-th/0108057, Phys. Lett. B, in press.
\bibitem{Sto94} V.\ G.\ J.\ Stoks, R.\ A.\ M.\ Klomp, C.\ P.\ F.\ Terheggen, 
and J.\ J.\ de Swart, Phys.\ Rev.\ C {\bf 49}, 2950 (1994).
\bibitem{WSS95} R.\ B.\ Wiringa, V.\ G.\ J.\ Stoks, and R.\ Schiavilla, 
Phys.\ Rev.\ C {\bf 51}, 38 (1995).
\bibitem{data99} The 1999 world $np$ database is the Nijmegen 1992
$np$ database~\cite{Sto93} plus all $np$ data published between January 1993
and December 1999 and not rejected in the analysis.
The latter data are listed in Table~XVI of Ref.~\cite{Mac01}.
\bibitem{43} J. E. McAninch, W. Haeberli, H. Wita{\l}a, W. Gl\"ockle, 
J. Golak, 
Phys.\ Lett.\ B {\bf 153}, 29 (1985).
\bibitem{44} S. Shimizu, K. Sagara, H. Nakamura, K. Maeda, T. Miwa, 
N. Nishimori, S. Ueno, T. Nakashima, and S. Morinobu, 
Phys.\ Rev.\ C {\bf 52}, 1193 (1995).
\bibitem{45} W. Tornow, C. R. Howell, R. C. Byrd, R. S. Pedroni, 
R. L. Walter, 
Phys.\ Rev.\ Lett.\ {\bf 49}, 312 (1982).
\bibitem{46} F. Sperisen, W. Gr\"uebler, V. Koenig, P. A. Schmelzbach, 
K. Elsener, B. Jenny, C. Schweizer, J. Ulbricht, P. Doleschall,  
Nucl.\ Phys.\ A {\bf 422}, 81 (1984).
\bibitem{47} H. R\"uhl, B. Dechant, J. Krug, W. L\"ubcke, G. Spangardt, 
M. Steinke, M. Stephan, D. Kamke, J. Balewski, K. Bodek, L. Jarczyk, 
A. Strza{\l}kowski, W. Hajdas, St. Kistryn, R. M\"uller, J. Lang, R. Henneck, 
H. Wita{\l}a, Th. Cornelius, W. Gl\"ockle, 
Nucl.\ Phys.\ A {\bf 524}, 377 (1991).
\bibitem{48} H. Shimizu, K. Imai, N. Tamura, K. Nisimura, K. Hatanaka, 
T. Saito, Y. Koike, Y. Taniguchi, 
Nucl.\ Phys.\ A {\bf 382}, 242 (1982).
\bibitem{49} K. Sagara, H. Ogari, S. Shimizu, K. Maeda, H. Nakamura, 
T. Nakashima, S. Morinobu, 
Phys.\ Rev.\ C {\bf 50}, 576 (1994).
\bibitem{50} H. Wita{\l}a, W. Gl\"ockle, L. E. Antonuk, J. Arvieux, 
D. Bachelier, B. Bonin, A. Boudard, J. M. Cameron, H. W. Fielding, 
M. Garcon, F. Jourdan, C. Lapointe, W. J. McDonald, J. Pasos, G. Roy, 
I. The, J. Tinslay, W. Tornow, J. Yonnet, W. Ziegler, 
Few-Body\ Systems\ {\bf 15}, 67 (1993).
\bibitem{51} G. Rauprich, S. Lemaitre, P. Niessen, K. R. Nyga, 
R. Reckenfelderb\"aumer, L. Sydow, H. Paetz gen. Schieck, H. Wita{\l}a, 
W. Gl\"ockle, 
Nucl.\ Phys.\ A {\bf 535}, 313 (1991).
\bibitem{52} H. R. Setze, C. R. Howell, W. Tornow, R. T. Braun, W. Gl\"ockle, 
A. H. Hussein, J. M. Lambert, G. Mertens, C. D. Roper, F. Salinas, 
I. Slaus, D. E. Gonz\'alez Trotter, B. Vlahovi\'c, R. L. Walter, 
H. Wita{\l}a, 
Phys.\ Lett.\ B {\bf 388}, 229 (1996).
\bibitem{53} J. Strate, K. Geissd\"orfer, R. Lin, J. Cub, E. Finckh, 
K. Gebhardt, S. Schindler, H. Wita{\l}a, W. Gl\"ockle, Th. Cornelius, 
J.\ Phys.\ G:\ Nucl.\ Phys.\  {\bf 14}, L 229 (1988).
\end{references}
\end{document}